\documentstyle[preprint,aps,epsfig,floats]{revtex}
\input epsf.tex
\def\DESepsf(#1 width #2){\epsfxsize=#2 \epsfbox{#1}}
\begin{document}
\preprint{\vbox{\hbox{}}}
\draft
\title{
CP Violation in Hyperon Decays}
\author{Xiao-Gang He
\footnote{Plenary talk presented at the Kaon2001,
International Workshop on CP violation, 
Pisa Italy, 12-17, June, 2001
}}
\address{
Department of Physics, National Taiwan University,
Taipei, Taiwan}
%\date{}

\maketitle

\begin{abstract}
In this talk I review theoretical predictions
for CP violation in non-leptonic hyperon decays in the Standard Model 
and models beyond. In the Standard Model the CP violating 
observable $A$ in the polarization asymmetries of $\Lambda \to p \pi^-$
and $\Xi^- \to \Lambda \pi^-$
decays are predicted to be in the ranges $(-0.61\sim 6.8)\times 10^{-5}$
and $-(0.1\sim 1)\times 10^{-5}$, respectively. These ranges are below the 
sensitivity of $1.4\times 10^{-4}$ for $A(\Lambda^0_-)+A(\Xi^0_-)$ 
for E871 experiment at Fermilab. When going beyond the SM, such as,
Supersymmetric and Left-Right symmetric models, 
$A$ can be as large  as $10^{-3}$ and a few times of $10^{-4}$, respectively.  
Studies of hyperon 
decays can provide important information about CP violation.
\end{abstract}
\baselineskip=14pt
\newpage
\section{CP Violating Observables in Hyperon decays}

The decay amplitude $M$ for a non-leptonic hyperon decay 
$B_i\to B_f \pi$ can be written as

\begin{eqnarray}
M = G_F m^2_\pi \bar B_f(q_f) (\tilde A-\tilde B\gamma_5) B_i(q_i).
\nonumber
\end{eqnarray}
In the literature one often uses 
$S=\tilde A$ and $P = \tilde B |\vec q_f|/(E_{f}+m_{f})$.

The polarization asymmetry interest for CP violation study 
is related to the polarization parameter $\alpha$
defined as follow,
\begin{eqnarray}
{d\Gamma \over d\Omega} = {\Gamma\over 4\pi}(1
+\alpha \hat q_f\cdot \vec \omega_i),\nonumber 
\end{eqnarray}
where $\vec\omega_i$ is the initial hyperon polarization direction
and $\hat q_f$ is the final baryon momentum direction,
and
$\alpha = 2Re(S^* P)/(|S|^2+|P|^2)$.

If particle and anti-particle decays are measured, 
one can construct a CP violating
observable\cite{1},

\begin{eqnarray}
A(B_i\to B_f \pi) 
= {\alpha + \bar \alpha\over \alpha -  \bar \alpha}.
\end{eqnarray}
In the CP conserving limit, $\alpha = -\bar \alpha$, and
therefore $A =0$.

So far only
upper bounds 
of $A(\Lambda^0_-)$ and $A(\Xi^-_-)$ for $\Lambda \to p \pi^-$ and
$\Xi^-\to \Lambda \pi^-$ have been obtained\cite{2}.
The experiment E871 at Fermilab underway can measure $A(\Lambda \to p \pi^-) + 
A(\Xi^-\to \Lambda \pi^-)$ with a sensitivity of $1.4\times 
10^{-4}$\cite{3}. With this sensitivity, it is possible to distinguish
several different models. Therefore important information
for CP violation can be obtained. 

To the leading order $A(\Lambda^0_-)$ and $A(\Xi^-_-)$ 
are given by the 
following\cite{2},

\begin{eqnarray}
A(\Lambda^0_-) 
\approx -\tan (\delta^p_1 - \delta_1^s) \sin(\phi^p_1 - \phi^s_1),
\;\;A(\Xi^-_-) 
\approx -\tan (\delta^p_2 - \delta_2^s) \sin(\phi^p_2 - \phi^s_2),
\nonumber
\end{eqnarray}
where $\delta^{s,p}_{1,2}$ and $\phi^{s,p}_{1,2}$ are the CP conserving
final state interaction phases and CP violating weak interaction phases of 
the $\Delta I = 1/2$ isospin amplitudes,
respectively.
$\delta^{s,p}_1$ extracted from data are: $\delta_1^s = 6.0^\circ$ and
$\delta^p_1 = -1.1^\circ$ with errors of order $1^\circ$\cite{4}. There is no
experimental information on $\delta^{s,p}_2$. Theoretical calculations
give $\delta^s_2 = 0.5^\circ$ and $\delta^p_2 = -1.7^\circ$\cite{5}.
We have

\begin{eqnarray}
A(\Lambda^0_-) \approx 0.125\sin(\phi_1^p-\phi_1^s),\;\;\;\;
A(\Xi^0_-) \approx 0.04 \sin(\phi^p_2-\phi^s_2).
\end{eqnarray}
To finally obtain $A$ one needs to calculate the phases $\phi_{1,2}^{s,p}$.
In order to minimize uncertainties, we use experimental data for the
CP conserving amplitudes and calculate the CP violating 
amplitudes to obtain these phases.

\section{Standard Model Predictions}

The CP violating decay amplitude $ImM$ is given by

\begin{eqnarray}
Im M &=& Im(<\pi B_f| H_{eff}|B_i>) = {G_F\over \sqrt{2}}V_{us}V_{ud}^*
Im(\tau) 
\sum_i <B_f \pi |y_i O_i|B_i>.\nonumber
\end{eqnarray}
Here $H_{eff}$ is the effective $\Delta S =-1$ Hamiltonian. 
$\tau =- V_{td}^*V_{ts}/V_{ud}^*V_{us}$, and 
$O_i$ are operators composed of quarks and gluons up to dimension six. 

Calculations show that the dominant 
contribution is coming from
$O_6 =\bar d_i \gamma_\mu(1-\gamma_5)s_j 
\sum _{q'=u,d,s} \bar q'_j \gamma^\mu(1+\gamma_5) q'_i$.
We will use results obtained in Ref.\cite{1} 
MIT bag model as the reference 
values. We have

\begin{eqnarray}
&&\phi^s_1 \approx 0.42 y_6 Im\tau B_{\Lambda s}^6,
\;\;\;\;\phi_1^p \approx 2.24 y_6 
Im \tau B_{\Lambda p}^6;\nonumber\\
&&\phi^s_2 \approx 0.29y_6 Im\tau B^6_{\Xi s}, 
\;\;\;\;\phi^p_2 \approx -0.92y_6 Im \tau B_{\Xi p}^6,
\nonumber
\end{eqnarray}
where $y_6=-0.0995$.
In the above, we have introduced the parameters $B^6_i$ to quantify the
uncertainty in these matrix elements with $B^6_i=1$ for bag model calculations.
To reflect uncertainties due to our poor understanding of the hadronic 
matrix elements, in our numerical
analysis we will, conservatively, use $0.5 < B^i_{j s} < 2$ and allow
$B^i_{j p}$ to vary in the range of $0.7 B^i_{js} < B^i_{j p} < 
1.3 B^i_{j s}$.

Here we also give bag model calculations for $O_{11}= 
{g_s\over 8 \pi^2} \bar d \sigma_{\mu\nu} G^{\mu\nu} (1+\gamma_5) s$ which 
will be important for our discussions on new physics.  
We have\cite{6}

\begin{eqnarray}
&&\phi^s_1 \approx 0.13 y_{11} Im\tau B^{11}_{\Lambda s},
\;\;\;\;\phi_1^p \approx 0.15 y_{11} B^{11}_{\Lambda p} 
Im \tau;\nonumber\\
&&\phi^s_2 \approx 0.08y_{11} Im\tau B^{11}_{\Xi s}, 
\;\;\;\;\phi^p_2 \approx -0.04y_{11} Im \tau B^{11}_{\Xi p}.
\nonumber
\end{eqnarray}
In the SM $y_{11} = -0.34$. 

Combining above information, we have

\begin{eqnarray}
&&A(\Lambda^0_-) = [0.28(B^6_{\Lambda p} - 0.19B^6_{\Lambda s})
y_6 + 0.019(B^{11}_{\Lambda p} - 0.87 B^{11}_{\Lambda s}) y_{11}]
Im(\tau), \nonumber\\
&&A(\Xi^-_-) = -[0.037(B^6_{\Xi p} + 0.32B^6_{\Xi s})y_6
+0.0016(B^{11}_{\Xi p} + 2.0 B^{11}_{\Xi s})y_{11}]Im(\tau).\nonumber
\end{eqnarray}

Using the expression $Im(\tau) = -A^2\lambda^4\eta$ with 
$\lambda = 0.2196$, 
$A = 0.835$\cite{4}, the best fit value $\eta= 0.34$\cite{7} and
$B_j^{6,11} =1$, 
we obtain

\begin{eqnarray}
A(\Lambda^0_-) = 1.2\times 10^{-5},\;\;\;\;A(\Xi^-_-) = -0.27\times 10^{-5}.
\end{eqnarray}
Other model 
calculations give similar values\cite{8}.

Using the 95\% C.L. allowed range of $0.22\sim 0.50$ for $\eta$ and the
conservative allowed ranges for $B^6_{i, (s,p)}$, 
we obtain

\begin{eqnarray}
A(\Lambda^0_-) = 
(-0.61\sim 6.8)\times 10^{-5},\;\;\;A(\Xi^-_-) = -(0.1\sim 1.0)\times 10^{-5}.
\end{eqnarray}

We consider the above the most conservative allowed ranges for the SM 
predictions. $|A(\Lambda^0_-)|$ is larger than $|A(\Xi^-_-)$. 
This is a general feature in
all models. From now on we will only discuss
$A(\Lambda^0_-)$.

\section{ Beyond the Standard Model}

When going beyond the SM, $A$ can be larger.
Model independent analysis shows that $A(\Lambda^0_-)$ as 
large as $10^{-3}$ is
possible\cite{8}. Here we consider two typical models, Supersymmetric and 
Left-Right symmetric models, to demonstrate that it is indeed possible to have
a large $A$. 

\subsection{
Supersymmetric Model}

In a general supersymmetric model there are more CP violating phases 
and new operators.
Among them gluonic dipole operators with new CP violating phase 
due to exchange of gluino and 
squark with left-right mixing in the loop can produce a $A(\Lambda^0_-)$
considerably larger than the SM prediction.
This new interaction has been shown to make large contribution to 
$\epsilon'/\epsilon_K$\cite{9} also. 

The effective Hamiltonian for the gluonic dipole
operator is\cite{10},
\begin{eqnarray} 
{\cal H}_{\it eff} &=& C_{g} 
{g_s\over 8\pi^2} m_s \bar d \sigma_{\mu\nu} G^{\mu\nu}
(1+\gamma_5) s + 
\tilde{C}_{g} 
{g_s\over 8\pi^2} m_s \bar d \sigma_{\mu\nu} G^{\mu\nu}
(1-\gamma_5) s ~+~{\rm h.c.},\nonumber
\nonumber
\label{effh}
\end{eqnarray}
where
\begin{equation}
C_{g} = (\delta^d_{12})_{LR} 
{\alpha_s \pi \over  m_{\tilde g} m_s} 
G_0(x) \;,\;
\tilde{C}_{g} = (\delta^d_{12})_{RL} 
{\alpha_s \pi \over  m_{\tilde g} m_s} 
G_0(x) .\nonumber
\label{wilco}
\end{equation}
The parameters $\delta_{12}^{d}$ characterize the mixing in 
the mass insertion approximation, and 
$x=m_{\tilde g}^2/m_{\tilde q}^2$, 
with $m_{\tilde g}$, $m_{\tilde q}$ being the gluino and averaged  
squark masses, respectively. The loop function can be found in Ref.\cite{10}. 

Using our previous MIT bag model results, we obtain\cite{11}
\begin{eqnarray}
        \lefteqn{
        A(\Lambda^0_-)_{SUSY} =
        \left({\alpha_s(m_{\tilde{g}}) \over \alpha_s(500~{\rm GeV})}
        \right)^{23\over 21}
        \left({500~{\rm GeV}\over m_{\tilde{g}}}\right) 
        {G_0(x)\over G_0(1)} 
        } \nonumber \\
        &\times&
        \left((2.0 B_p - 1.7 B_s) {\rm Im}(\delta_{12}^d)_{LR}
        +(2.0 B_p + 1.7 B_s) {\rm Im}(\delta_{12}^d)_{RL} \right).
        \nonumber
\label{cpasym}
\end{eqnarray}

\begin{figure}[t]
  \vspace{5.5cm}
  \includegraphics{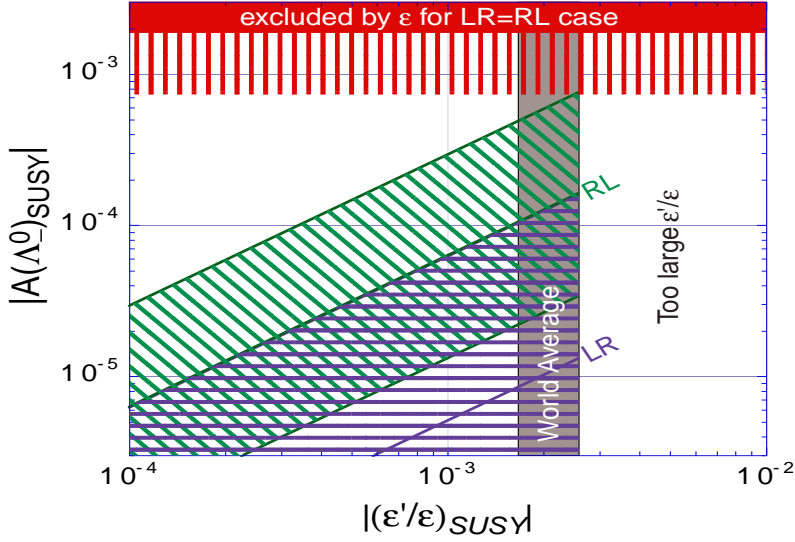}
\vspace{1cm}
  \caption{\it
Allowed region for $|A(\Lambda)|$.    
\label{figure} }
\end{figure}

There are constraints from $\epsilon_K$ and $\epsilon'/\epsilon_K$.
$\epsilon'/\epsilon_K$ constrains the linear combination of
$Im((\delta_{12}^d)_{LR} - (\delta^d_{12})_{RL})$ while
$\epsilon_K$ constrains $Im((\delta^d_{12})_{LR} + (\delta^d_{12})_{RL})$.
We consider three cases with results shown in Fig.\ref{figure}:
a) ${\rm Im}(\delta_{12}^d)_{RL}=0$ (hatched horizontally),
b) ${\rm Im}(\delta_{12}^{d})_{LR}=0$ (hatched diagonally), and 
c) ${\rm Im}(\delta_{12}^d)_{RL}$ = $ 
{\rm Im}(\delta_{12}^d)_{LR}$ (below the shaded region or
vertically hatched), for illustrations.
The regions allowed, by the requirements that $(\epsilon'/\epsilon_K)_{SUSY}$ 
and $(\epsilon_K)_{SUSY}$ not to exceed their experimental values, for 
the three cases discussed are shown in 
Fig.~\ref{figure}.  
It is clear that the SUSY contribution can be very different than that in the
Standard Model. $A(\Lambda^0_-)$ can be as large as $1.9\times 10^{-3}$.

\subsection{Left-Right Symmetric Model}

In Left-Right symmetric model, there are two charged gauge bosons,
$W_L$ and $W_R$. In general
there are mixing between these two bosons. 
The effective Hamiltonian for non-leptonic hyperon decays can be written
as

\begin{eqnarray}
H_{eff} = H_{SM} + H_{R} + H_{LR}.\nonumber
\end{eqnarray}
In the limit of no left-right mixing, 
$H_{SM}$ reduces to the SM effective Hamiltonian.
$H_{R}$ is due to the exchange of the heavy $W$ boson which can
be obtained from $H_{SM}$
by replacing $m_W = m_{W_1}$ by the heavy boson mass $m_{W_2}$, 
the left-handed KM matrix $V_L$ by the right-handed KM matrix 
$V_R$, and $1\pm \gamma_5$ by $1\mp \gamma_5$.
$H_{LR}$ is due to a non-zero parameter $\xi$ for the left-right mixing. 
It is given by

\begin{eqnarray}
H_{LR} &=& {G_F\over \sqrt{2}}
{g_R\over g_L} \xi
[ V_{Lud}^*V_{Rus} \bar d \gamma^\mu (1-\gamma_5) u \bar u
\gamma_\mu(1+\gamma_5) s \nonumber\\
&+& V_{Rus}^*V_{Lus}\bar d \gamma^\mu (1+\gamma_5)
u \bar u \gamma_\mu(1-\gamma_5) s\nonumber\\
&+& \sum_i \tilde G(x_i){g_s\over 8\pi^2}
m_i \bar d \sigma^{\mu\nu} G_{\mu\nu}
(V_{Rid}^*V_{Lis}(1-\gamma_5) + V^*_{Lid}V_{Tis}(1+\gamma_5) )s],\nonumber
\end{eqnarray}
where $\tilde G(x) = -3x\ln x/2(1-x)^3 - (4+x+x^2)/4(1-x)^2$ and 
$x_i = m_i^2/m_W^2$.

The contributions from $H_R$ is suppressed by a factor of $m^2_W/m^2_{W_2}
<10^{-3}$
compared with the SM ones if the elements in 
$V_{L}$ and $V_{R}$ are similar in 
magnitudes. 
However, $H_{LR}$ can have large contributions even though there is a
suppression factor $\xi < 4\times 10^{-3}$\cite{12}. The reason is that 
there is a chiral enhancement factor $m^2_K/(m_u+m_d)(m_s-m_u)$ 
for the four quark operators in $H_{RL}$. Also there is an enhancement
factor $m_t/m_s$ for the gluonic dipole operator compared with the
SM one.

For the four quark operator contributions in $H_{LR}$, with QCD corrections,
one obtains\cite{13}

\begin{eqnarray}
\phi^s_1 = -12.3 \xi^u_-,\;\;\;\;\phi^p_1 = -0.33 \xi^u_+,
\
\end{eqnarray}
where $\xi^i_\pm = \xi Im(V_{Lid}^*V_{Ris}\pm V_{Rid}^*V_{Ris})$.
Using $\xi = 4\times 10^{-3}$, $A(\Lambda^0_-)$ can be as large as
$10^{-4}$ if $Im(V_{Rud}^*V_{us})$ is larger than 0.1 which is not ruled.

The contributions from the gluonic dipole interactions can be obtained in a
similar way as in the SUSY case. Including QCD corrections we 
obtain\cite{13}

\begin{eqnarray}
\phi_1^s = -0.54\sum_i \xi^i_- \tilde G(x_i) {m_i\over \mbox{GeV}},\;\;\;\;
\phi^p_1 = 0.63\sum_i\xi^i_+ \tilde G(x_i) {m_i\over \mbox{GeV}}.
\end{eqnarray}
The above contributions are similar to that considered in the previous section
for the SUSY contributions. 
However there are some differences that the parameters $\xi$ and the elements
of $V_{L,R}$ are constrained from other experimental data and unitarity of 
$V_{L,R}$ which are sever than constraint from $\epsilon_K$. 
$A(\Lambda^0_-)$ can not be as large as that in the SUSY case.
With $|V_{L td}| =0.003$ and $V_{R ts} \approx 0.04$, one obtains a
$A(\Lambda^0_-)$ in the order of a $\mbox{few}\times 10^{-4}$. 

\section{Conclusions}

In the Standard Model the CP violating 
observables $A(\Lambda, \Xi)$ 
in the polarization asymmetries of $\Lambda \to p \pi^-$
and $\Xi^-\to \Lambda \pi^-$
decays are predicted to be in the ranges $(-0.61\sim 6.8)\times 10^{-5}$ 
and $-(0.1\sim 1.0)\times 10^{-5}$. These asymmetries are 
below the sensitivity of
$1.4\times 10^{-4}$ of E871 at Fermilab 
experiment. When going beyond the SM, such as,
Supersymmetric and Left-Right symmetric models, 
$A$ can be as large as $10^{-3}$ and a few times of $10^{-4}$, respectively.
Studies of hyperon 
decays can provide important information about CP violation.

This work was supported in part by NSC of R.O.C. under grant number
NSC89-2112-M-002-058 and by NCTS. 
I thank D. Chang, J. Donoghue, H. Murayama, S. Pakvasa, H. Steger 
and G. Valencia 
for collaborations on materials presented in this paper.

\end{document}